\documentclass[runningheads]{llncs}
\usepackage[T1]{fontenc}
\usepackage{comment}
\usepackage{graphicx}
\usepackage{longtable}
\usepackage[
 svgnames,
 table,
]{xcolor}
\usepackage[
 colorlinks,
 urlcolor = NavyBlue,
 citecolor = NavyBlue,
 linkcolor = NavyBlue,
]{hyperref}
\usepackage{lscape}
\usepackage{array}
\usepackage{booktabs}
\usepackage{amsmath}
\usepackage{amssymb}
\usepackage{algorithm}
\usepackage{algpseudocode}
\usepackage{float}  
\usepackage{paralist}
\usepackage{orcidlink}
\let\orcidID\orcidlink
\usepackage{todonotes}
\usepackage{xspace}
\usepackage{framed}
\usepackage{martinmacros-stripped}
\usepackage{rotating}  

\let\oldcite\cite

\renewcommand{\cite}[1]{{\small\oldcite{#1}}}

\newcommand{\Obs}{\mathit{O}} %
\newcommand{\domainP}[1]{\mathcal{D}(#1)} %

\newcommand{\movesto}[1]{\stackrel{#1}{\rightarrow}}

\newcommand{\forget}[1]{}  %

\newcommand{\shortVersion}[1]{}

\newcommand{\alphabet}{\Sigma}
\newcommand{\transf}{\delta}

\newcommand{\tuple}[1]{( #1 )}

\newcommand{\auto}{\cA\xspace} %
\newcommand{\sizeOf}[1]{|#1|} %

\newcommand{\langP}[1]{\mathcal{L}(#1)}

\newcommand{\hide}[1]{}

 \graphicspath{{images}}

\algdef{SE}[SUB]{Subroutine}{EndSubroutine}[2]{\textbf{Subroutine} \textsc{#1}(#2)}{\textbf{End Subroutine}}

\begin{document}

\title{Inference of Deterministic Finite Automata\\ via Q-Learning}

\author{%
Elaheh Hosseinkhani%
\orcidID{0000-0002-3410-8808} \and
Martin Leucker%
\orcidID{0000-0002-3696-9222}
}

\institute{
Universit{\"a}t zu L{\"u}beck, L{\"u}beck, Germany\newline\email{$\{$Elaheh.Hosseinkhani,leucker$\}$@isp.uni-luebeck.de} 
}
\maketitle 
\begin{center}
\textcolor{red}{\textbf{This is a preprint of a paper that will appear in SBMF 2025.}}
\end{center}

\begin{abstract}

Traditional approaches to  inference of deterministic finite-state automata (DFA) stem from symbolic AI, including both active learning methods (e.g., Angluin’s L* algorithm and its variants) and passive techniques (e.g., Biermann and Feldman’s method, RPNI). Meanwhile, sub-symbolic AI, particularly machine learning, offers alternative paradigms for learning from data, such as supervised, unsupervised, and reinforcement learning (RL). This paper investigates the use of \emph{Q-learning}, a well-known reinforcement learning algorithm, for the passive inference of deterministic finite automata. It builds on the core insight that the learned Q-function, which maps state-action pairs to rewards, can be reinterpreted as the transition function of a DFA over a finite domain. 
This provides a novel bridge between sub-symbolic learning and symbolic representations.
The paper demonstrates how Q-learning can be adapted for automaton inference and provides an evaluation on several examples. 
\end{abstract}

\begin{keywords}
  Automata Learning,    
  Q-Learning,     
  Reinforcement Learning,
  Symbolic and Sub-symbolic AI Integration
\end{keywords}

\section{Introduction}

Many interactive systems can be effectively modeled as automata, allowing for structured analysis and understanding of their behavior. Beyond system modeling, automata have broad applications in fields such as software verification, natural language processing, and network protocol analysis. As a result, automata learning--the process of automatically constructing automata from observations--is a compelling area of research with practical relevance across multiple domains. 

One of the most commonly used and well-studied machine models is that of deterministic finite-state automata 
(DFA)--and we focus on DFA also in this paper. The problem of learning such automata is addressed within the field of grammatical inference, which has received considerable attention in recent years. There are two principle ways to learn an automaton: learn an automaton from a black-box system itself, i.e., \emph{active learning}, or learn an automaton directly from observational data, i.e., \emph{passive learning.} The active learning setting has been the most studied in the literature~\cite{Angluin87,Vaandrager17}. In that setting, the process of learning an automaton consists of querying the black-box system with two types of queries: \emph{membership queries}, to check if a specific sequence of observational data is accepted by the system, and \emph{equivalence queries}, to check if the language accepted by the automaton corresponds to the ones described by the system. A key research direction involves extending the learning process to more expressive classes of automata, like non-deterministic finite automata \cite{DBLP:conf/ijcai/BolligHKL09} and various classes of data automata \cite{DBLP:journals/corr/BolligHLM14,DBLP:conf/apn/DeckerHLT14}. Many of the algorithms for active learning elaborate on the Nerode's right congruence classes which characterize a minimal DFA uniquely. To this end, a characteristic set that uniquely identifies the automaton is eventually obtained \cite{BergGJLRS05} by queries and storing the results in a clever way. Roughly, a characteristic set consists of strings leading to states as well as strings distinguishing states.

Active learning has several limitations. For instance, it may require many interactions with the black-box system, which can be prohibitively costly. This drawback is exacerbated by the fact that the equivalence queries can often be done only statistically and therefore may entail heavy testing, if one requires a high accuracy. In addition, the numerous interactions also suppose a potentially unlimited access to the black-box system, which in real-world situations would often not be possible. Another drawback of active learning is that it completely leaves out all the available observational data, a resource that is often abundantly available.\footnote{Admittedly, existing observational data, e.g.\ previously recorded traces in form of log files can easily be integrated in the initial phase of typical active learning algorithms like Angluin's L$^*$\cite{Angluin87}. } By contrast, passive learning is data-driven: an automaton is learned directly from the data without interaction with the black-box system.

Until now, the study of automata passive learning dealt with DFA (deterministic finite automaton) and NFA (non-deterministic finite automaton), and several techniques have been used to handle this learning problem. Since learning (minimal) automata is NP-complete in general~\cite{DBLP:journals/iandc/Gold67}, a popular method is constraint-solving, which consists in encoding the passive learning problem as a set of constraints over integers \cite{DBLP:journals/tc/BiermannF72}. A solution over the numbers 1 to $k$ exists iff there is a corresponding automaton with $k$ states. Such constraint problem may be translated into a satisfiability problem of propositional logic, which can be solved using a SAT solver \cite{DBLP:conf/cade/GrinchteinLP06}. 
Regular Positive and Negative Inference (RPNI) \cite{RPNI} is a further classic algorithm for learning deterministic finite automata (DFA) from labeled examples. It starts by building a prefix tree acceptor  
and then merges states while ensuring consistency with both positive and negative examples. RPNI is efficient in practice but in general, no minimal automaton is obtained.

All mentioned learning algorithms are part of so-called symbolic AI, the field that, in simple words, uses explicit symbols and rules to represent knowledge and logic and corresponding algorithms. Sub-symbolic AI, on the other hand, refers to AI approaches that do not use explicit, symbolic representations of knowledge but often often numerical methods to process and learn from data. Especially sub-symbolic AI has gained a lot of attention in recent years and our main research question in this paper is to find out whether one of the techniques of sub-symbolic AI is usable for learning automata. 

In general, one may distinguish three different kinds of subsymbolic learning techniques:  \emph{Supervised Learning}, which relies on labeled data, where each input has a corresponding correct output. The model, often different forms of neural networks, learns to map inputs to outputs based on these explicit examples. \emph{Unsupervised Learning} deals with unlabeled data, aiming to find patterns, structures, or relationships within the data without predefined outcomes. In \emph{Reinforcement Learning (RL)}, an \emph{agent} learns to make sequential decisions by interacting with an \emph{environment.} The agent receives \emph{rewards} or \emph{penalties} for its actions, and its goal is to maximize the cumulative reward over time. 

One of the first works using sub-symbolic techniques for (passively) learning an automaton was \cite{Aichernig2024}, which used supervised learning for estimating a recurrent neural network which is subsequently turned into an automaton. In this paper, we investigate whether reinforcement learning can be used for obtaining DFAs. More precisely, we 
use Q-learning for learning a strategy eventually representing the transition function of the automaton in question. 

As a prototypical model-free RL, Q-Learning \cite{russell2021artificial} enables agents to learn through trial and error, without requiring explicit knowledge of the environment. The agent explores a final set of \emph{states} that represent different configurations of the environment. A corresponding set of actions defines the options available in each state. Upon taking an action, the agent receives feedback in the form of a reward. Over time, it gradually improves its decisions using a process known as \emph{temporal-difference learning}.

For automata inference (see Section~\ref{sec:qlearning_for_automata}), we adapt the Q-function as a symbolic structure encoding transitions over a finite input domain. More precisely, the states in the sense of Q-learning are automaton states plus input letter and the action that the agent may choose and which is optimized is the automaton successor state for the given state and input letter--the transition. A further difficulty arises as, potentially, each (automaton) successor state may be an accepting (final) or a non-accepting(non-final) state. 

In summary, we address the question of whether Q-learning can be applied to passive automata learning. We answer this question positively by introducing the algorithm Q-PAI which stands for Q-learning-based passive automata interference. We give a detailed comparison to the work in \cite{Aichernig2024} and to RPNI, showing that Q-learning often outperforms the other approaches on the considered benchmarks. This paper is organized as follows: Section~\ref{sec:preliminaries} introduces the necessary preliminaries, including fundamental concepts of automata, automata learning, the motivation for our approach, and an overview of Q-learning. Section~\ref{sec:qlearning_for_automata} presents our method for using Q-learning to infer automata. Section~\ref{sec:evaluation} reports the evaluation, including the experimental setup and results. Section \ref{sec:conclusion} concludes the paper and outlines directions for future work. Additional algorithms are provided in Appendix~\ref{sec:appendix_algorithm}.

\section{Preliminaries}
\label{sec:preliminaries}

\subsection{Automata}

Let $\N$ denote the natural numbers, and, for $n \in \N$, let $[n] :=
\{ 1, \dots, n\}$. For the rest of this section, we fix an alphabet
$\Sigma$.
A \Def{deterministic finite automaton} (DFA) $\auto 
= \tuple{S, s_0,
  \transf,S^+}$ over $\alphabet$ consists of a finite set of
\Def{states} $S$, an \Def{initial state} $s_0 \in S$, a
\Def{transition function} $\transf : S \times \alphabet \to S$, and a
set $S^+ \subseteq S$ of \Def{accepting states}.
A \Def{run} of $\auto$ is a sequence $s_0 \movesto{\sigma_1} s_1
\movesto{\sigma_2} \dots \movesto{\sigma_n} s_n$ such that $\sigma_i \in \alphabet$, $s_i \in S$ and $\transf(s_{i-1}, \sigma_i) = s_i$ for all $i \in [n]$. It is called \Def{accepting} iff $s_n \in S^+$. The \Def{language}
accepted by $\auto$, denoted by $\langP{\auto}$, is the set of strings
$u \in \alphabet^\ast$ for which an accepting run exists.  Since the
automaton is deterministic, it is reasonable to call the states
$S\setminus S^+$ also \Def{rejecting states}, denoted by $S^-$. We
extend $\transf$ to strings as usual by $\transf(s, \epsilon) = s$ and
$\transf(s, u\sigma) = \transf((\transf,u),\sigma)$. The size of $\auto$,
denoted by $\sizeOf{\auto}$, is the number of its states $S$, denoted
by $\sizeOf{S}$. A language is \Def{regular} iff it is accepted by
some DFA. A language $L$ is called \Def{prefix closed} iff for all $u\sigma \in L$ also $u\in L$. We call an automaton prefix closed iff its accepted language is prefix closed.

\subsection{Automata Learning}

A \emph{sample} is a set of strings that, by the language in question,
should either be accepted, denoted by $+$, or rejected, denoted by
$-$. For technical reasons, it is convenient to work with
prefix-closed samples.  As the samples given to us are not necessarily
prefix closed we introduce the value \Def{maybe}, denoted by $?$.
Formally, a \Def{sample} is a partial function $\Obs : \alphabet^\ast
\to \{ + , -, ?\}$ with finite, prefix-closed domain $\domainP{\Obs}$.
That is, $\Obs(u)$ is defined only for finitely many $u \in
\alphabet^\ast$ and is defined for $u \in \alphabet^\ast$ whenever it
is defined for some $ua$, for $a \in \alphabet$.  For a string $u$ the
sample $\Obs$ yields whether $u$ should be \emph{accepted},
\emph{rejected}, or we do not know (or do not care).  For strings $u$
and $u'$, we say that $\Obs$ \emph{disagrees} on $u$ and $u'$ if
$\Obs(u)\neq ?$, $\Obs(u')\neq ?$, and $\Obs(u)\neq \Obs(u')$.
Sometimes it is convenient to identify $\Obs$ with a (finite) set of observations $\Obs = \{(u,+) \mid \Obs(u) = + \} \cup \{(u,-) \mid \Obs(u) = - \}$ and the context identifies when we consider $\Obs$ as  function or as a set. An automaton $\cA$ is said to \Def{conform} with a sample $\Obs$, if
whenever $\Obs$ is defined for $u$ we have $\Obs(u) = +$ implies $u
\in \langP{\cA}$ and $\Obs(u) = -$ implies $u \notin \langP{\cA}$.

We are now ready to define the problem studies in this paper:

\begin{definition}[Passive Learning Problem]
Given a sample $\Obs$, the concept of \Def{passive learning} is the task to obtain a minimal DFA $\auto$ that conforms with $\Obs$. 
\end{definition}

\subsection{Q-Learning}
\label{sec:qlearning}

 \paragraph{Reinforcement Learning (RL)}~\cite{Robotica_1999} is a framework for sequential decision-making in which an agent interacts with an environment to learn a policy that maximizes cumulative reward. The learning process is driven by state transitions and delayed feedback, without requiring an explicit model of the environment. In contrast to supervised learning, RL does not rely on labeled input-output pairs but instead discovers behavior through exploration and reward signals. This generality makes RL particularly appealing for tasks in which the system's structure is unknown or only partially observable.

 \paragraph{Q-Learning} is a foundational model-free RL algorithm. It assumes that a finite set of \emph{states} $Q$ is given, representing possible states of the environment the agent may be in. The set of \emph{actions} $A$ denotes the possible actions the agent may pick in each state. A \emph{reward function} $R : Q \times A \to \mathbb{R}$ provides the direct award when choosing an action $a$ in a state $q$. In the Q-learning algorithm an agent incrementally constructs a value function \( Q : Q \times A \rightarrow \mathbb{R} \) that estimates the utility of executing action \( a \in A \) in state \( q \in Q \). The update rule adjusts \( Q \) based on observed rewards ($R$) and the estimated value of successor states, following the principle of temporal difference learning. Notably, Q-Learning is off-policy, enabling the agent to evaluate the optimal policy independently of the current exploration strategy. In the context of automata inference (see Section~\ref{sec:qlearning_for_automata}), we reinterpret the Q-function as a symbolic structure encoding transitions over a finite input domain, thereby connecting statistical learning with the formal construction of DFA \cite{russell2021artificial}. The generic form of Q-learning is defined in Algorithm~\ref{alg:q-learning}. 

 Once the Q-table is initialized arbitrarily (line~1), the Q-function is updated (line~8) a fixed number of times (line~2). Hereby, one may update either the values for all $q$ or only for those for reaching a certain goal (line~4). The update of the Q-function is weighted based on the so-called \emph{learning rate} $\alpha \in (0,1]$ and the \emph{discount factor} for future rewards $\gamma \in [0,1]$. For choosing the next action $a$ to explore (line~5), serveral strategies may be chosen, some of which depend on a so-called exploration rate $\epsilon \in [0,1]$.

{
\footnotesize
{\begin{algorithm}[t]

\caption{Generic Q-Learning Procedure\label{alg:q-learning}}

\begin{algorithmic}[1]
\State Initialize the Q-table $\mathcal{Q}(q, a)$  for all $q \in Q$, $a \in A$
\For{each episode}
    \State Select a random initial state $q$
    \While{goal not reached}
        \State Choose action $a$ using an exploration strategy (e.g., $\epsilon$-greedy)
        \State Execute action $a$
        \State Observe reward $r\gets R(q,a)$ and next Q-learning State $q'$
        \State Update Q-function:
       {\scriptsize $ \mathcal{Q}(q, a) \leftarrow \mathcal{Q}(q, a) + \alpha \left[ r + \gamma \cdot \max_{a' \in A} \mathcal{Q}(q', a') - \mathcal{Q}(q, a) \right]
        $}
        \State Set $q \leftarrow q'$
    \EndWhile
\EndFor
\end{algorithmic}
\end{algorithm}}}
\paragraph{Strategy} Our learning strategy is based on trajectory exploration followed by reward-based Q-value updates. For each word, the agent builds a path through abstract states using either exploration (random transitions) or exploitation (choosing the best-known transition). The Q-table is updated twice: once based on the label of the terminal state, and once based on whether the constructed DFA accepts the word correctly. This double-reward structure helps the agent learn transitions that are both locally and globally correct.
\section{Using Q-Learning for Learning Automata}
\label{sec:qlearning_for_automata}
\normalsize

We are now ready to present our Q-learning-based approach for passively inferring a DFA from a given sample $O$. As described in Section~\ref{sec:qlearning}, Q-learning operates on states and actions, estimating the best action to take in a given state. In our context, the goal is to learn the transitions of a DFA--that is, to determine the successor state given a current automaton state and the input letter to be read. Simply put, we aim to learn, for each automaton state and input symbol, which state the transition should lead to. As such, the states in the sense of Q-learning are pairs of automaton states and input letters. The actions in the sense of Q-learning are the (automaton) states. Notably, each potential successor state appears in two variants: accepting (final) and non-accepting (non-final). Therefore, the Q-learning function must estimate both the target state and whether it should be accepting or rejecting.

To prevent confusion between the Q-learning process and the resulting DFA, we explicitly distinguish between their respective representations. Consider we are estimating a DFA $\cA = (S, s_o, \delta, S^+)$ over the alphabet $\Sigma$. Then we define the learning space as follows: 
${Q} = S \times \Sigma$ is the set of states for Q-learning, $F = \{ +,- \}$ indicates automata states to be final ($+$) or non-final ($-$),  $A = S \times F$ is the set of actions the Q-learning algorithm may choose from. Each Q-value is defined via the function $\mathcal{Q} : Q \times A \to \mathbb{R}$ which is in fact a function 
$\mathcal{Q} : S \times \Sigma \times S \times F \rightarrow \mathbb{R}$. The value $\mathcal{Q}(s, \sigma, s',f) \in \mathbb{R}$ represents the learned utility of taking transition $s \xrightarrow{\sigma} s'$, where $s'$ has acceptance status $f$. We organize the learned Q-values in a structured Q-table, with row labels ranging over pairs of automata states and input labels and column labels ranging over the automata states paired with $+$ and $-$. Then, the Q-table  maps each transition—defined by a state, input symbol, next state, and acceptance status—to a corresponding utility value. Note that the Q-table has shape of $n \times |\Sigma|$ rows and $2n$ columns for a DFA of $n$ states in question.
{
\footnotesize
\begin{algorithm}[t]
\caption{ Q-PAI – Q-learning-based Passive Automata Inference\label{alg:q-learning-automata}}
\begin{algorithmic}[1]
\Require \textbf{Global Constants } $\alpha$, $\gamma$, $\epsilon_{min}$, $r$
\Function{Q-PAI}{$ \Sigma, O, n, \mathcal{E}$}
\State Initialize Q-tables $\mathcal{Q}$, $\mathcal {Q}^*\gets$ zero matrices of dimension $(n \cdot |\Sigma|) \times (2n)$
\State Initialize DFAs $\cA$ and $\cA^*$ with initial state $s_0$ 
\For{episode $= 1$ to $\mathcal{E}$}
    \For{ each sample $(w=\sigma_1\ldots\sigma_k, y) \in O$ } 
        \State Set $s \gets s_0$, $\tau \gets \left[~\right]$, $\chi \gets  0$
        \While{$\chi = 0$}
            \For{$i = 1$ to $|w|$}
                \State $ q \gets (s,\sigma_i)$
                \State $a = (s',f) \gets$ \Call{ExploreOrExploit}{$\mathcal{Q}$, $q$, $i$}
                \State Append $(s, \sigma_i, s',f)$ to $\tau$; set $s \gets s'$
            \EndFor
            \State $\mathcal{Q}$, $\mathcal{A}$, $\chi$, $\mathcal{Q^*}$, $\mathcal{A^*} \gets$ \Call{EvaluateAndUpdate}{$\mathcal{Q}$, $\mathcal{Q^*}$, $O$, $\tau$, $w$, $y$, $\chi$}
            \If{$\chi = 2$}
                \State \Return Inferred DFA $\mathcal{A}^*$
            \EndIf
        \EndWhile
    \EndFor
\EndFor
\State \Return Inferred DFA $\mathcal{A}^*$
\EndFunction
\end{algorithmic}
\end{algorithm}
}
\normalsize
The general Q-learning algorithm (Algorithm~\ref{alg:q-learning}) is then instantiated to a specialized version for passively learning automata resulting in Algorithm~\ref{alg:q-learning-automata}. Here, we use the notation summarized in Table~\ref{tab:notationQLearning}.
\begin{table}[!ht]
    \caption{Notation for Q-Learning%
    \label{tab:notationQLearning}}%
    {
\begin{framed}
{\footnotesize
\begin{itemize}
    \item[$k$:] The index of the current symbol in the input word. This value is often used to modulate exploration intensity (e.g., more exploration early in the word, more exploitation later).
    \item[$\epsilon$:] The current exploration probability---the likelihood of choosing a random action rather than the best known one.
    \item[$\epsilon_{\min}$:] A minimum bound on $\epsilon$, used to ensure that the agent never completely stops exploring.
    \item[$\alpha$:] The \textbf{learning rate}, determining how much new information overrides the existing Q-values. A typical value lies in $(0,1)$.
    \item[$\gamma$:] The \textbf{discount factor}, specifying how much future rewards are taken into account. When $\gamma = 0$, the agent only considers immediate rewards; when $\gamma \to 1$, it prioritizes long-term rewards.
    \item[$\tau$:] The \textbf{transition trajectory}, a list of state-action-state triples $(s_i, \sigma_k, s_j)$ recorded during the processing of a sample word $w$. It captures the agent’s path through abstract learning states $\mathcal{Q}$.
    \item[$O$:] The \textbf{observation }, a finite set of labeled strings: $O = \{ (w_1, y_1), \ldots, (w_n, y_n) \}$, where $w_i \in \Sigma^*$ and $y_i \in \{+, -\}$ indicates acceptance by the target language. 
    \item[$\mathcal{Q}$:] The \textbf{Q-table}, a function $\mathcal{Q} : {S} \times \Sigma \times S\times F \rightarrow \mathbb{R}$ storing utility values for transitions in the learning space.
    \item[$\mathcal{Q}^*$:] The \textbf{best Q-table} encountered during training, based on DFA accuracy.
    \item[$\mathcal{A}$:] The \textbf{current DFA} constructed from the current Q-table via a greedy policy.
    \item[$\mathcal{A}^*$:] The \textbf{best DFA} discovered so far during training, according to evaluation on data smaples $O$.
    \item[$\chi$:] An integer flag that controls processing of the current word: $\chi=0$ means reprocess the word, $\chi=1$ means skip it, and $\chi=2$ terminates the Q-learning procedure.%
\end{itemize}%
}
\end{framed}%
}
\end{table}

Let us explain the algorithm in detail: While in the general Q-learning algorithms, each update of the Q-table improves it, this is not the case here, as we explain below. Therefore, we operate with two Q-tables $\cQ$ and $\cQ^*$ where $\cQ$ is our working table while $\cQ^*$ keeps the currently best table. Moreover, we work with both, the Q-table as well as a direct representation of it as an automaton. The Q-table is used for updating Q-values while the automaton representation is used to evaluate the Q-table in terms of an automaton. Again, we work with two automata representations $\cA$ and $\cA^*$ with an initial state $s_0$, again using $\cA$ for temporary updates while keeping the so-far best version in $\cA^*$. In lines 2 and 3, we initialize the two Q-tables as well as the automata. We learn for $\mathcal{E}$ episodes, line 4. Each episode consists of processing each sample, line 5, letter by letter (line 8), building-up a path from the initial state $s_0$ (line 6) and choosing the successor state according to the  Q-table (line 10) and appending it to the current trace $\tau$ (line 11).
The selection of the action to choose next (i.e., the successor state of the automaton to explore) is determined via function \textsc{ExploreOrExploit} (line 10). When the whole word $w$ has been processed, the Q-tables and automata are reevaluated and updated based on the whole trace $\tau$ via function \textsc{EvaluateAndUpdate}. Depending on the reevaluation, the word $w$ is processed again ($\chi=0$) or the next word in the sample is considered ($\chi=1$). 
Eventually, all samples have been processed $\cE$ times and the algorithm returns the learned automaton $\cA^*$ (line 20). If during learning an automaton is found that conforms to the observed behavior, the goal of the algorithm is reached and the resulting %
automaton $\cA^*$ is returned (line~15). 

Let us now turn our attention to the two subroutines \textsc{ExploreOrExploit} and \textsc{EvaluateAndUpdate}, which explore the best action (successor state) and update the Q-tables/automata, respectively. 

\paragraph{Adaptive Exploration Based on Q-Value Variance.}
Let us start with explaining \textsc{ExploreOrExploit}($\mathcal{Q}$, $q$, $i$), %
see Algorithm~\ref{alg:exploreAndExploit}. It takes the current Q-table, the current state (automaton state and letter $\sigma$), and the current position in the word considered.
Its goal is to either pick the best action according to the current Q-table (line 7) or to pick one randomly, line 5. The latter is done to address the issue of premature convergence to suboptimal policies. We implemented an adaptive exploration strategy that dynamically adjusts the exploration probability based on the statistical variance of Q-values in a given state. This mechanism allows the agent to explore more in uncertain or under-trained regions of the state-action space. 
The core idea is to compute an adaptive $\epsilon$ from the variance of Q-values at the current state (lines 2,3). A higher variance suggests greater uncertainty in action quality and thus warrants more exploration. Concretely, the dynamic threshold builds on the variance of the values in the row of $q$ (line 2), amplified by the position $i$ of the letter to consider and compared to 1.0 and $\epsilon_\mathrm{min}$, lines 2-3. 
The choice of whether to explore randomly or deterministically depends on randomly choosing a value and comparing it to the dynamic threshold $\epsilon$, line 4. 

{
\footnotesize
\begin{algorithm}[t]
\caption{Explore and Exploit Subroutine\label{alg:exploreAndExploit}}
\begin{algorithmic}[1]
\Subroutine{ExploreAndExploit}{$\mathcal{Q}, q=(s,\sigma), i$}
    \State $\mathcal{V} \gets \text{Var}(\mathcal{Q}[s, \sigma, \cdot, \cdot])$ 
    \State $\epsilon \gets \max(\epsilon_{\min}, \min(1.0, i \cdot \mathcal{V}))$
    
    \If{$\text{rand()} < \epsilon$}
        \State Randomly sample $(s', f)$ from$\mathcal{Q}[s, \sigma, \cdot, \cdot]$
    \Else
        \State $(s', f) \gets \arg\max_{(s'', f')} \mathcal{Q}[s, \sigma, s'', f']$
    \EndIf

    \State \Return $(s', f)$
\EndSubroutine
\end{algorithmic}
\end{algorithm}
}
{\footnotesize
\begin{algorithm}[t]
\caption{\textsc{Evaluate and Update Subroutine}\label{alg:evalAndUpdate}}
\begin{algorithmic}[1]
\Subroutine{EvaluateAndUpdate}{$\mathcal{Q}, \mathcal{Q}^*,O,\tau, w, y, \chi$ }
    \State $(\_,\_,\_, \hat{y_1}) \gets$ last element of $\tau$ 
    \State $r_1 \gets$ \Call{ComputeRewardFromQtable}{$\hat{y_1}, y$ }
    \State$ \mathcal{Q} \gets$ \Call{UpdateQTable}{$\mathcal{Q}, \tau, r_1$}

    \State $\mathcal{A} \gets$ \Call{DFAFromQ}{$\mathcal{Q}$}
    \State $\hat{y}_2 \gets \mathcal{A}(w)$
   
    \State $r_2 \gets$ \Call{ComputeRewardfromDFA}{$\hat{y}_2$, $y$}
    \State $\mathcal{Q} \gets$ \Call{UpdateQTable}{$\mathcal{Q}, \tau, r_2$}
    \State $\mathcal{A} \gets$ \Call{DFAFromQ}{$\mathcal{Q}$}
    \State $\hat{y}_2 \gets \mathcal{A}(w)$
    \State $\chi \gets 1 $ \textbf{if }{ $\hat{y}_2 = y $}
     \If{$\text{Accuracy}(\mathcal{A}, O) = 1$}
        \State $\mathcal{Q}^* \gets \mathcal{Q}$; \quad $\mathcal{A}^* \gets \mathcal{A}$; $\chi \gets 2$
               
    \Else
    
        \If{\Call{Accuracy}{$\mathcal{A}, O$} $>$ \Call{Accuracy}{$\mathcal{A}^*, O$}}
            \State $\mathcal{Q}^* \gets \mathcal{Q}$; $\mathcal{A}^* \gets \mathcal{A}$ %
        \EndIf
    \EndIf

    \State \Return $\mathcal{Q}$, $\mathcal{A}$, $\chi$, $\mathcal{Q}^*$, $\mathcal{A}^*$
\EndSubroutine
\end{algorithmic}
\end{algorithm}
}

\normalsize
\paragraph{Evalutate and Updating the Q-table} Let us now turn our attention towards the function \textsc{EvaluateAndUpdate}($\cQ$, $\cQ^*$, $O$, $\tau$, $w$, $y$, $\chi$), see Algorithm~\ref{alg:evalAndUpdate}.
The main task of this function is to update the Q-table. This is less obvious as in the general Q-learning algorithm (Algorithm~\ref{alg:q-learning}) as the Q-table holds two versions of each automaton state, once as final state, once as non-final state. Clearly, any automaton has each state either as final or non-final state. Therefore, we roughly do the following: We compute rewards according to the Q-table and update it correspondingly (lines 3,4). Moreover, we translate the Q-table into an automaton (line~5), hereby choosing each automaton state to be final or non-final. We also use the resulting automaton to compute a reward and use it for updating the Q-table (line 8). 

Thus, the Q-table is updated twice, once by general means according to Q-learning, once by interpreting the table as an automaton. If this results in an automaton that conforms to the sample, we are done (lines 12,13). If not, we check whether the current automaton $\cA$ fits better to the sample than the previously best automaton $\cA^*$. If so, we take it, otherwise we ignore the update of the Q-table and the resulting automaton. Moreover, if the classification of the word $w$ under consideration as final/non-final state compared with that in the automaton does not match, the word is re-considered for further processing.

\begin{algorithm}[H]
\caption{\textsc{Reward Computation from DFA Subroutine}}
\label{alg:rewardfromdfa}
\begin{algorithmic}[1]
\Subroutine{ComputeRewardFromDFA}{$ \hat{y}, y$}
    \If{$\hat{y} = y$}
        \State $R \gets r$
    \Else
        \State $R \gets  {-r/2}$
    \EndIf
    \State \Return $R$
\EndSubroutine
\end{algorithmic}
\end{algorithm}

\begin{algorithm}[H]
\caption{\textsc{Reward Computation from Q-table Subroutines}}\label{alg:rewardfromQtable}
\begin{algorithmic}[1]
\Subroutine{ComputeRewardFromQtable}{$ \hat{y}, y$}
    \If{$y = \hat{y}$}
        \If{$\mathit{Label} = 0$}
            \State $R \gets 2 \times r$
        \Else
            \State $R \gets 4 \times r$
        \EndIf
    \Else
        \State $R \gets {-r/2}$
    \EndIf
    \State \Return $R$
\EndSubroutine
\end{algorithmic}
\end{algorithm}

\paragraph{Q-table update.} The Q-table is updated exactly as in the general case for Q-learning (Algorithm~\ref{alg:q-learning}, line~8), applied for each letter in the considered state sequence $\tau$, resulting in Algorithm~\ref{alg:updateQtable}.
{
\footnotesize
\begin{algorithm}[t]
 \caption{\textsc{Update QTable Subroutine} \label{alg:updateQtable}}
 \begin{algorithmic}[1]
   \Subroutine{Update Q-Table}{$\mathcal{Q},\tau ,\ r$}
   \For{each $(s, \sigma, s', f) \in \tau$}

      \State {\footnotesize $\mathcal{Q}[s, \sigma, s', f] \gets \mathcal{Q}[s, \sigma, s', f]$ 
            \par 
      \hskip\algorithmicindent 
      \hskip\algorithmicindent 
      \hskip\algorithmicindent 
      \hskip\algorithmicindent 
      \hskip\algorithmicindent 
      \hskip\algorithmicindent 
        $+ \ \alpha \cdot \left[ r + \gamma \cdot \max\limits_{(s'', f') \in \mathcal{S} \times F} \mathcal{Q}[s, \sigma, s'', f'] - \mathcal{Q}[s, \sigma, s', f] \right]$
      }
   \EndFor
   \State \Return $\mathcal{Q}$
   \EndSubroutine
  \end{algorithmic}
\end{algorithm}
}
{
\footnotesize
 \begin{algorithm}[!ht]
\caption{\textsc{DFA from Q Subroutine}\label{alg:DFAfromQ}}
    \begin{algorithmic}[1]
        \Subroutine{DFAFromQ}{$\mathcal{Q},  O, \Sigma$}
        \State Initialize $\pi^* \gets \{\}$, $\delta \gets \{\}$, $\mathcal{S} \gets \{s_0\}$, $\mathcal{S}^+ \gets \{\}$
        
        \For{each state $s$ and symbol $\sigma$ in $\mathcal{Q}$}
            \State $\pi^*[s, \sigma] \gets \arg\max_{(s', f)} \mathcal{Q}[s, \sigma, s', f]$
        \EndFor
        
        \For{each $(w, y) \in O$} 
            \If{$y = +$}
                \State $s \gets s_0$
                \For{each symbol $\sigma_i$ in $w$}
                    \State $(s', f) \gets \pi^*[s, \sigma_i]$
                    \State $\delta(s, \sigma_i) \gets s'$
                    \State $\mathcal{S} \gets \mathcal{S} \cup \{s'\}$
                    \State $s \gets s'$
                    \If{$f = +$}
                        \State $\mathcal{S}^+ \gets \mathcal{S}^+ \cup \{s'\}$
                    \EndIf
                \EndFor
            \EndIf
        \EndFor

        \State $\mathcal{A} \gets $\Call{CompleteDFAwithSink}{$\mathcal{S}, s_0, \delta, S^+$}
        \State \Return $\mathcal{A}$
        \EndSubroutine
    \end{algorithmic}
\end{algorithm}

}
\normalsize
\paragraph{DFA Construction from Q-table.} Given a learned Q-table $\mathcal{Q}$, we aim to construct a deterministic finite automaton (DFA) that captures the policy embedded in $\mathcal{Q}$. This process involves three main steps: extracting optimal actions, pruning transitions, and identifying final states (see Algorithm~\ref{alg:DFAfromQ}). First, lines 2-4, we compute the optimal policy $\pi^*$ by selecting the action $(s', f)$ that maximizes the Q-value for each state-symbol pair $(s, \sigma)$, formally:
$\pi^*[s, \sigma] = \arg\max_{(s', f)} \mathcal{Q}[s, \sigma, s', f].$
Next, we prune the state-action graph using only the positively labeled trajectories in the observation set $O^+$. For each accepted word $w = \sigma_1 \ldots \sigma_k$, we simulate the path from the initial state $s_0$ by following the optimal actions in $\pi^*$. The corresponding transitions $\delta(s, \sigma_i) = s'$ are added, and the set of visited states is updated accordingly, lines~6--13.
During this process, if the target label $f$ of any optimal action is $+$, the target state $s'$ is added to the final state set $\mathcal{S}^+$ (line 15). This ensures that the DFA accurately reflects the terminal predictions encoded in the Q-table. Finally, we construct the DFA as a tuple: $\mathcal{A} = (\mathcal{S}, s_0, \delta, \mathcal{S}^+)$
 where $\mathcal{S}$ is the set of all visited states, $s_0$ is the initial state, $\delta$ is the transition function, and $\mathcal{S^+}$ is the set of final (accepting) states.

\normalsize
\paragraph{DFA Completion with Sink State.} To ensure that the DFA that \(\delta\) is total for all \(s \in \mathcal{S}, a \in \Sigma\), we introduce a \emph{Sink} state ($\bot$) in Algorithm~\ref{alg:CompSink} in Appendix~\ref{sec:appendix_algorithm} that adds ($\bot$) if not already present, redirects all undefined transitions $(s,\sigma)$ to $(\bot)$ and append $\delta(\bot,\sigma)\gets \bot$ for all \(\sigma \in \Sigma\).
\paragraph{${Accuracy}$ $(\mathcal{A}, O).$} 
The accuracy of an automaton $\mathcal{A}$ with respect to a given sample $O$ is defined as the proportion of words in $O$ that are classified correctly (as accepted or non-accepted) by $\mathcal{A}$. Formally:
\[\scalebox{0.95}{$
\mathrm{Accuracy}(\mathcal{A}, O) \;=\; 
\frac{\left|\{(w, y) \in O \mid \mathcal{A}(w) = y\}\right|}{|O|}$}
\]
where $\mathcal{A}(w)$ denotes the classification of $w$ by $\mathcal{A}$.
The computation procedure is given in Algorithm ~\ref{alg:accuracy} in Appendix ~\ref{sec:appendix_algorithm}.
\textit{Reward Computation.}
The reward is computed by checking whether the suggested state (final/non-final) coincides with the samples classification, giving more weight to positive labels. Moreover, the main reward is given for the Q-table, while less reward is applied  when the DFA is considered. The exact numerical value have been tuned empirically based on the experimental section. Formally, let $y \in \{+, -\}$ be the sample label 
and $\hat{y}$ the predicted label for a given word. Then:
\[
\scalebox{0.80}{$
R_{\mathrm{DFA}}(y, \hat{y}) =
\begin{cases}
r & \text{if } \hat{y} = y \ \land\ y= + \\
0 & \text{if } \hat{y} = y \ \land\ y= - \\
-\dfrac{r}{2} & \text{otherwise}
\end{cases}
\quad
R_{\mathrm{Qtable}}(y, \hat{y}) =
\begin{cases}
4r & \text{if } \hat{y} = y \ \land\ y = + \\
2r & \text{if } \hat{y} = y \ \land\ y = - \\
-\dfrac{r}{2} & \text{otherwise}
\end{cases}
$}
\]
\normalsize
Here, the  $R_\mathrm{Qtable}$ corresponds to the reward assigned after processing an entire word. In this, a correct classification of a negative label yields $2r$, correct classification of a positive label yields $4r$, and any mismatch results in a penalty of $-r/2$. $R_\mathrm{{DFA}}$  corresponds to the reward assigned after verifying the DFA’s final/non-final state prediction for the current word, where a reward of $r$ is given only when the prediction matches a positive label $(y = +)$, a penalty of $-r/2$ is applied for any mismatch, and zero reward is given for correctly classified negative labels.  The exact numerical values of these weights ($r$, $2r$, $4r$, $-r/2$) have been tuned empirically, as described in the experimental section.

This concludes the description of our inference algorithm Q-PAI. Putting a bound on the while-loop, our algorithm will always terminate. It is formally unclear, whether the algorithm will provide a conforming automaton if one exists. That said, if the algorithm does not terminate with a conforming example, one might re-execute Q-PAI with a larger $n$ for the target automaton size. Note that a conforming automaton always exists with $n$ matching the size of the sample. As in many applications of reinforcement learning, we do not have strong formal guarantees of the algorithm but in practice it performs well. So, we continue with the evaluation of Q-PAI evaluation in the next section.

\section{Evaluation}
\label{sec:evaluation}
\normalsize
\paragraph{Data Sets.} 
To evaluate the applicability of Q-learning for automata inference, we employed two types of benchmark datasets: Tomita grammars \cite{tomita1982} and the BLE communication traces collected by \cite{Aichernig2024} which represent real-world data from the Bluetooth Low Energy protocol. The Tomita grammars are a standard benchmark of seven regular languages over \{0,1\}, each defined by a minimal DFA with varying complexity. As a practical case study, we use BLE traces from real devices containing three system on the chips (SoCs) CYBLE-416045-02 (3 states), nRF52832 (5 states), CC2650 (5 states), each representing symbolic input sequences of full protocol sessions.  In this dataset the whole input alphabet was considered.
For each dataset, we employ three types of training sets to evaluate learning performance under different input conditions:
\begin{inparaenum}[(1)]
\item Characteristic set, minimal set of samples sufficient for identifying the target DFA. 
\item AAL-Generated Data, samples generated via the AALpy framework~\cite{AALPY}, simulating active learning interactions such as membership and equivalence queries; and 
\item Random Data, samples drawn randomly from the input space to reflect unstructured or naturalistic behavior. 
\end{inparaenum}
These datasets pose greater challenges due to larger alphabets and behavioral variability, allowing us to evaluate the scalability and robustness of the approach.

During training, we make no assumptions about the total number of states. The initial number of states is set to $|w|_{\min}$, the length of the shortest word in the sample set. If no automaton can be found that correctly matches the observations, the number of states is incremented by one, and the procedure is repeated. Our experiments aim to answer the following core questions: 
\begin{inparaenum}[(1)]
    \item Can Q-learning, in a passive setting, recover the correct structure of the target DFA given sufficiently informative samples?
    \item How does the learned automaton behave under variations in the input data, such as characteristic sets, AAL-generated samples, and randomly sampled inputs?
    \item How does the method compare to existing symbolic learners with respect to accuracy, robustness, and sample efficiency?
\end{inparaenum}

\paragraph{Experimental Setup.}
All experiments were conducted on a MacBook Pro (Model Identifier: \texttt{Mac14,7}) equipped with an Apple M2 chip featuring 8 cores (4 performance and 4 efficiency), 16 GB unified memory, and macOS. The experiments were executed in a Python 3.10 environment using a custom implementation of the Q-learning framework. Each Q-learning model was trained over a maximum of $\mathcal{E} = 200$ episodes, with adaptive exploration controlled by a variance-based $\epsilon$-strategy (see Algorithm~\ref{alg:exploreAndExploit}). We applied a fixed learning rate range $\alpha = 0.1 $ and discount factor $\gamma = 0.9$.  Through empirical tuning, we observed that a reward parameter of r = 1 led to improved convergence behavior and higher accuracy across our experiments. For each grammar, the learner is trained on labeled strings and evaluated on a disjoint test set. Metrics include learning accuracy, inferred automaton size, number of episodes and learning time. DFA extraction follows the convergence of the Q-table, after which pruning and final state determination are applied.

\paragraph{Results.}
Table~\ref{tab:evaluationtomita} and Table~\ref{tab:evaluationBLE} highlight the comparative analysis of Q-Learning, RNN, and RPNI across a variety of automata inference tasks. Q-learning effectively inferred minimal DFAs with high accuracy, particularly when characteristic datasets were provided. In most Tomita grammars, Q-learning achieved $100\%$ accuracy with minimal standard deviation. Even in more complex cases such as Tomita~5 and Tomita~7, accuracy remained high (e.g., $95.0 \pm 7.02\%$ and $89.0 \pm 4.14\%$, respectively), while still ensuring minimal state representations. These results confirm that Q-learning consistently identifies the correct number of DFA states, as expected from the target languages, and exhibits reliable convergence behavior. Q-learning also demonstrated stable and efficient runtime characteristics. Training durations remained under one minute for simpler grammars and scaled reasonably for more complex cases (e.g., approximately $17$ minutes for Tomita~5 and $3.55$ minutes for Tomita~7). 

On BLE datasets, Q-learning maintained accuracy above $90\%$ in all scenarios, achieving $95.45 \pm 7.12\%$ and $90.05 \pm 2.87\%$ on `nRF52832` and $95.09 \pm 2.67\%$ on `cc2650`. Additionally, the inferred DFAs matched the expected sizes for each device, confirming the method’s robustness in real-world protocol modeling. Overall, Q-learning provides a reliable, scalable, and interpretable framework for automaton learning, delivering both minimality and accuracy without significant computational overhead. In comparison to other approaches, Q-learning consistently yielded more compact models than RNNs, which tend to overestimate the number of states. While RPNI is theoretically grounded, its performance was found to be highly sensitive to the dataset structure and often failed to produce minimal or accurate DFAs across complex or real-world inputs. These findings highlight Q-learning’s advantage in balancing interpretability, minimality, and predictive performance across both synthetic and practical domains. However, unlike formal methods such as RPNI, Q-learning lacks theoretical guarantees of convergence to the correct minimal DFA. Accuracy slightly declines in more complex scenarios--for instance, Tomita~5 and Tomita~7 achieve  $95.0 \pm 7.02\%$ and $89.0 \pm 4.14\%$ accuracy, respectively, indicating difficulty in generalizing to grammars with more intricate structure. Training time also increases substantially with dataset complexity, reaching up to $17$ minutes for Tomita~5 and over $47$ minutes for certain BLE datasets (e.g., \texttt{nRF52832}). In summary, our evaluation is promising though further evaluations are beneficial.

\begin{sidewaystable}[p] 
    \centering
    \begin{minipage}{0.95\textwidth} %
        \centering
        \caption{Evaluation results of Q-Learning, RNN, and RPNI on Tomita grammars.\vspace{-0.3cm}}
        \includegraphics[width=1\textwidth]{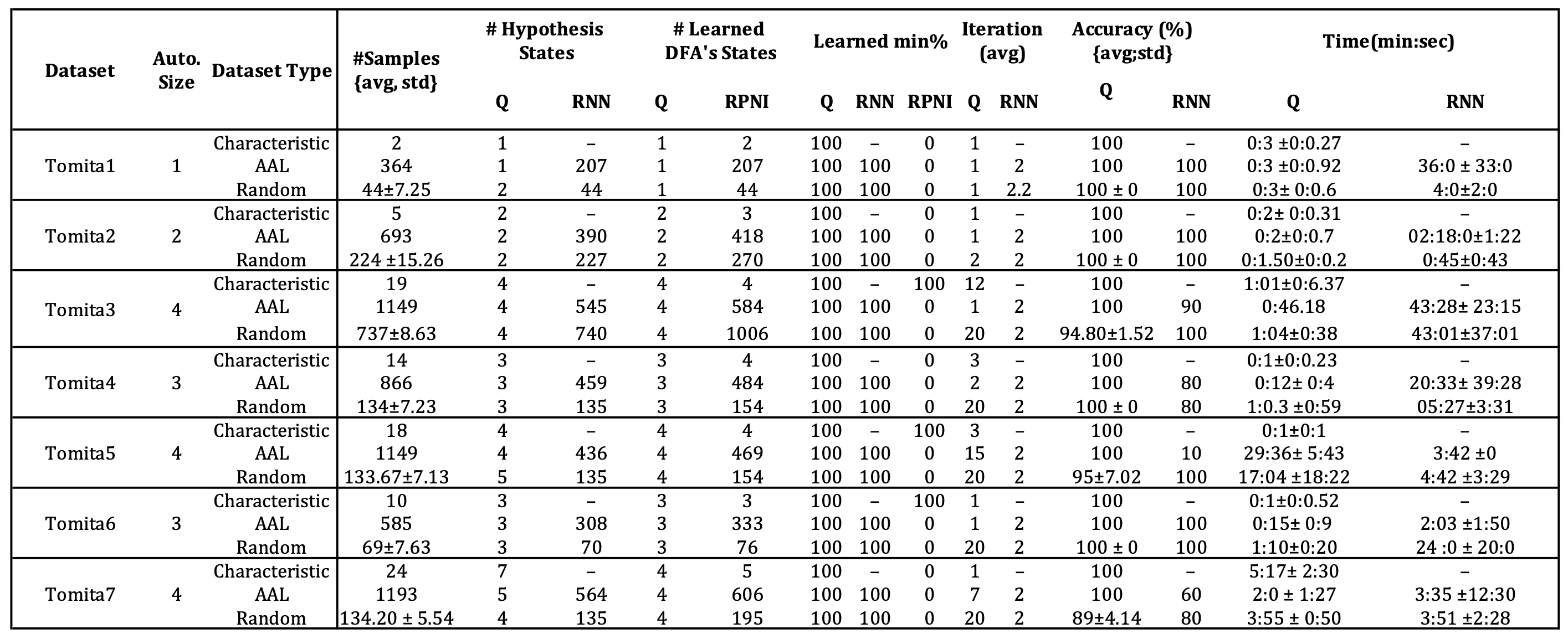}
        \label{tab:evaluationtomita}
    \end{minipage}
\\
    \vspace{-0.5cm}
    \begin{minipage}{0.95\textwidth} %
        \centering
        \caption{Evaluation results of Q-Learning, RNN, and RPNI on BLE devices.\label{tab:evaluationBLE}\vspace{-0.8cm}}
        \includegraphics[width=1\textwidth]{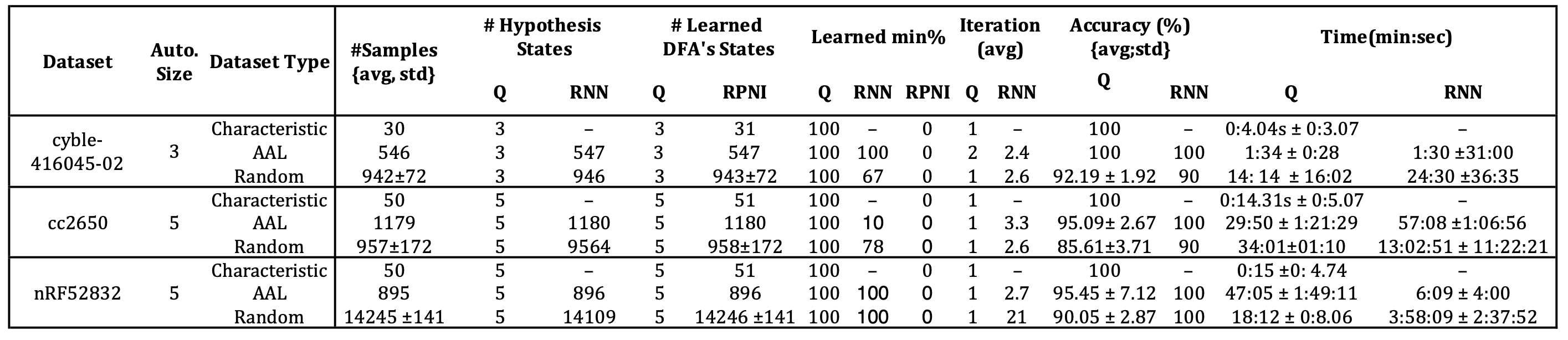}
        \end{minipage}
\end{sidewaystable}

\section{Conclusion}
\label{sec:conclusion}

This paper introduced Q-PAI, a Q-learning-based algorithm for passively inferring an automaton given a finite sample. 
With Q-learning being a sub-symbolic learning mechanism to derive symbolic structures, we  effectively link model-free learning with formal automata representations. Our approach demonstrates strong empirical performance in both accuracy and generalization. Its ability to infer compact, minimal DFAs across diverse dataset types, including characteristic sets, random traces, and AAL-generated samples—makes it a promising candidate for practical formal language learning tasks. 

However, this approach also presents several challenges. First, it is highly sensitive to the design of the reward function, which must be carefully tuned to reflect learning objectives and class balance. Second, the choice of exploration strategy can introduce inefficiencies in large or sparse state-action spaces, potentially leading to slower convergence. As the size of the target DFA increases, the scalability of the Q-table is expected to become a limiting factor due to higher memory and computational demands. Moreover, unlike formal algorithms such as RPNI, Q-learning lacks theoretical guarantees for convergence and minimality, relying instead on empirical performance and heuristic tuning. These limitations indicate that while Q-learning is effective and practical in many settings, future work should address its scalability, efficiency, and theoretical grounding to strengthen its applicability.

\bibliographystyle{splncs04}
\bibliography{bib}
\newpage
\appendix

\section{Additional Algorithms}\label{sec:appendix_algorithm}

 \begin{algorithm}[H]
    \caption{\textsc{Accuracy Subroutine}}\label{alg:accuracy}
    \begin{algorithmic}[1]
    \Subroutine{Accuracy}{$\mathcal{A}$, $O$}
    \State $correct \gets 0$
    \For{each $(w, y) \in O $}
        \State $\hat{y} \gets${$\mathcal{A}(w)$}
        \If{$\hat{y} = y$}
            \State $correct \gets correct + 1$
        \EndIf
    \EndFor
    \State $n \gets$ \Call{Length}{$O$}
    \State \Return $correct / n$
    \EndSubroutine
    \end{algorithmic}
\end{algorithm}

\begin{algorithm}[H]
    \caption{\textsc{Complete DFA with Sink}\label{alg:CompSink}}
    \begin{algorithmic}[1]
    \Subroutine {CompleteDFAwithSink} {$\mathcal{S}, s_0, \delta, S^+ $}
        \If {$\bot \notin \mathcal{S}$}
            \State $S \gets S \cup \bot$
        \EndIf
        \ForAll{ $\sigma \in \Sigma $}
        \ForAll{$ s \in \mathcal{S}$}

            \If{$\delta(s,\sigma) $ is undefined}
                \State $\delta(s,\sigma) \gets \bot$
            \EndIf
        \EndFor
        \State $\delta(\bot,\sigma) \gets \bot$
        \EndFor
    
    \State \Return $(\mathcal{S}, s_0, \delta, S^+)$
    \EndSubroutine
    \end{algorithmic}
\end{algorithm}

\end{document}